\documentclass[letterpaper]{proc}

\usepackage{algorithm2e}
\usepackage{amssymb}
\usepackage{amsmath}
\usepackage{graphicx}
\newcommand{\sgn}{\mathrm{sgn}}

\makeatletter
\def\@maketitle{%
  \vbox to 2.3in{%
    \hsize\textwidth
    \linewidth\hsize
    \vspace*{1.5cm}
    \centering
    {\bfseries\LARGE \@title \par}
    \vskip 2em
    {\large \begin{tabular}[t]{c}\@author \end{tabular}\par}
    \vfill}    \vspace*{1.0cm}
}
\renewcommand\section{\@startsection {section}{1}{\z@}%
     {.7\baselineskip plus\baselineskip}{.5\baselineskip}
                                   {\normalfont\Large\bfseries}}
\renewcommand\section{\@startsection {section}{1}{\z@}%
      {.5\baselineskip\@plus.7\baselineskip}{.3\baselineskip}%
                                   {\normalfont\Large\bfseries}}
\renewcommand\subsection{\@startsection{subsection}{2}{\z@}%
       {.5\baselineskip\@plus.7\baselineskip}{.3\baselineskip}%
                                   {\normalfont\large\bfseries}}
\renewcommand\subsubsection{\@startsection{subsubsection}{3}{\z@}%
      {.5\baselineskip\@plus.7\baselineskip}{.3\baselineskip}%
                                     {\normalfont\normalsize\bfseries}}
\makeatother

\renewenvironment{abstract}%
  {\normalfont
    \list{}{\labelwidth0pt
      \leftmargin0pt \rightmargin\leftmargin
      \listparindent\parindent \itemindent0pt
      \parsep0pt
      
    }%
    \item[\hskip\labelsep\bfseries\abstractname\enspace --] \itshape%
}{%
  \endlist}

\newcommand{\keywordsname}{Keywords}
\newenvironment{keywords}%
  {\normalfont
    \list{}{\labelwidth0pt
      \leftmargin0pt \rightmargin\leftmargin
      \listparindent\parindent \itemindent0pt
      \parsep0pt
      }%
    \item[\hskip\labelsep\bfseries\keywordsname:]}{\endlist}

\begin{document}

\title{Cooperative Training for Attribute-Distributed Data:\\
Trade-off Between Data Transmission and Performance\thanks{This
research was supported in part by the National Science Foundation
under Grant CNS-06-25637, and in part by the Office of Naval
Research under Grants N00014-09-1-0342, W911NF-07-1-0185 and
N00014-07-1-0555.}}

\author{{\bf Haipeng Zheng}\\
Dept. of Electrical Eng. \\
Princeton University\\
Princeton, NJ, U.S.A.\\
\underbar{haipengz@princeton.edu}\\
\and
{\bf Sanjeev R. Kulkarni}\\
Dept. of Electrical Eng.\\
Princeton University\\
Princeton, NJ, U.S.A.\\
\underbar{kulkarni@princeton.edu} \and
{\bf H. Vincent Poor}\\
Dept. of Electrical Eng.\\
Princeton University\\
Princeton, NJ, U.S.A.\\
\underbar{poor@princeton.edu}}
\date{}

\maketitle

\begin{abstract}
This paper introduces a modeling framework for distributed
regression with agents/experts observing attribute-distributed data
(heterogeneous data). Under this model, a new algorithm, the
iterative covariance optimization algorithm (ICOA), is designed to
reshape the covariance matrix of the training residuals of
individual agents so that the linear combination of the individual
estimators minimizes the ensemble training error. Moreover, a scheme
(Minimax Protection) is designed to provide a trade-off between the
number of data instances transmitted among the agents and the
performance of the ensemble estimator without undermining the
convergence of the algorithm. This scheme also provides an upper
bound (with high probability) on the test error of the ensemble
estimator. The efficacy of ICOA combined with Minimax Protection and
the comparison between the upper bound and actual performance are
both demonstrated by simulations.
\end{abstract}


\section{Introduction}
\noindent Distributed learning is a field that generalizes classical
machine learning algorithms to a distributed framework. Unlike the
classical learning framework, where one has full access to all the
data and has unlimited central computation capability, in the
framework of distributed learning, the data are distributed among a
number of agents, who have limited access to the data. These agents
are capable of exchanging certain types of information, which, due
to limited computational power and communication restrictions
(limited bandwidth or limited power), is usually restricted in terms
of content and amount. Research in distributed learning seeks
effective learning algorithms and theoretical limits within such
constraints on computation, communication, and confidentiality.

Distributed learning can be categorized into many subareas. In terms
of the way that the data are distributed, it can be categorized into
homogeneous data (instance/horizontally distributed data) and
heterogeneous data (attribute/vertically distributed data). In terms
of the structure of the entire system, it can be categorized into
systems with a fusion center and systems without a fusion center.
Each category has its unique applications and challenges. We focus,
in this paper, on the case in which data are attribute-distributed.
The algorithm that we develop can be adapted to systems either with
or without a fusion center.

The homogeneous-data problems have been widely studied. Two
important types of models are established in \cite{ref1} and
\cite{ref3} respectively: instance distributed learning with and
without a fusion center. The relationship between the information
transmitted among individual agents and the fusion center and the
ensemble learning capability are discussed in these papers.
Classical learning algorithms are more easily adapted to the
homogeneous cases because for each agent, the form of the
classifier/estimator is exactly the same as that of the centralized
learning algorithm. The homogeneity in the individual
classifiers/estimators is a great advantage for designing
distributed learning algorithms that compare and combine them.

However, these advantages disappear in the heterogeneous data case,
where different agents observe different attributes, and thus have
many different forms of classifiers/estimators. This makes it harder
to evaluate, compare and combine the estimators. Nevertheless, there
are some research results in this area (e.g., \cite{c1}, \cite{c3}).
Some basic ideas include voting/averaging, meta-learning, collective
data mining, and residual refitting. The voting/averaging algorithm
simply combines (linearly) the predictions of the individual agents.
The training process is purely non-cooperative. In the meta-learning
case (see \cite{ref2} and \cite{c6}), the fusion center seeks a more
sophisticated way to integrate predictions of individual estimators
by taking their predictions as a new training set (learning of
learning results), i.e., the fusion center treats the output of
individual estimators as the input covariates. Although this
hierarchical training scheme looks more delicate, it is still
non-cooperative and hence fails to learn hidden rules in which
covariates of different agents intertwine in a complicated way.

In contrast, the collective data mining algorithms (see \cite{c2},
\cite{c4} and \cite{c5}) are cooperative. They seek the information
required to be shared among the agents so that the optimal estimator
can be decomposed into an additive form without compromising the
performance of the ensemble estimator (compared to the estimator
trained by the centralized algorithm). Yet this requirement is
rather strong and hence this technique relies on specific types of
transformations, which require much prior knowledge of the problem,
and thus is hard to generalize to other problems. The residual
refitting algorithm, another cooperative training algorithm
described in \cite{c1}, has the advantage of not being dependent on
individual learning algorithms. The only way that the agents
communicate with each other is through their residuals. However,
these algorithms are based on an additive model, and are susceptible
to overtraining and pitfalls of local optima, though under some
assumptions, optimality can be guaranteed.

In this paper, we develop another cooperative training scheme, using
a modeling framework similar to that of the residual-refitting
algorithms. However, instead of refitting the residuals directly,
our new algorithm seeks to reshape the covariance matrix of the
residuals generated by all the agents so that the linear combination
of the estimators maintained by the agents can achieve a low
ensemble test error. Again, residuals are the only information that
the agents communicate to each other, yet they are used more
intelligently than in the case of residual-refitting. In the case
when residual-refitting is guaranteed to achieve global optimality,
our new algorithm, iterative covariance optimization algorithm
(ICOA) also achieves similar results - and due to its
insusceptibility to overtraining, ICOA usually outperforms.

Another important issue for a distributed learning system is the
trade-off between the amount of information exchanged among the
agents and the performance of the ensemble estimator/classifier. To
study this relationship, the major challenge is how to quantify the
information exchanged and the ensemble performance. In this paper,
based on ICOA with Minimax Protection, the relationship between the
amount of information exchanged (measured by the compression rate)
and the optimal test error of the ensemble estimator is discussed.
More interestingly, an upper bound on the test error of the ensemble
estimator is also derived.

The rest of this paper is organized as follow. In Section 2, we
describe the basic model and abstract the problem of finding an
optimal additive ensemble estimator into a two-stage optimization.
In Section 3, we analyze this optimization problem and introduce
ICOA, with its efficacy demonstrated by simulation. In Section 4, we
discuss the problem of how to keep ICOA functioning when the
covariance is not accurately estimated, which leads to Minimax
Protection, and we demonstrate the trade-off between data
transmission and system performance with an upper bound on the test
error with respect to the data compression rate. Section 5 contains
our conclusions.

\section{Model and problem}
\noindent
Our discussion is based on an estimation/regression problem
with attribute-distributed data. The estimation problem is specified
as follow:

There are $M$ covariates (or attributes) $X_1,\ldots,X_M$ and one
outcome $Y$, so the entire data set of $N$ instances is comprised of
\begin{displaymath}
\{(x_{i1},x_{i2}, \ldots, x_{iM},y_{i})\}_{i=1}^{N}
\end{displaymath}
where $N$ is the number of instances, $x_{ij}\in \mathbb{R}$ is the
$i$-th instance of $X_j$, and $y_{i}\in \mathbb{R}$ is the $i$-th
instance of $Y$.

We assume that there exists a hidden deterministic function
(rule/hypothesis) $\phi: \mathbb{R}^M \rightarrow \mathbb{R}$ such
that
\begin{equation}
y_i = \phi(x_{i1}, x_{i2},\ldots ,x_{iM}) + w_i,
\end{equation}
where $\{w_i\}_{i=1}^{N}$ is an independently drawn sample from a
zero-mean random variable $W$ that is independent of
$X_1,\ldots,X_M$ and $Y$.

Suppose there are $D$ agents, each of which has only limited access
to certain attributes. Define $F_j\text{~}(j=1,\ldots,D)$ to be the
set of attributes accessible by agent $j$, and define $F = \cup
_{j=1}^D F_j$, assuming that $|F|=M$. The outcome $Y$, with all its
instances, is visible to all the agents. These assumptions specify
the ``attribute-distributed" properties of our problem.

To highlight the distributed nature of the system, we add an extra
restriction: the only information that the agents can communicate
with each other is their training residuals (or information that can
be locally derived from the training residuals). This is a
reasonable assumption considering that the data observable by one
agent are usually confidential or incompatible with the learning
algorithm run by another agent.

Therefore, for these $D$ agents, each agent $i$ maintains an
estimator $f_i$ of the outcome, which is a function that takes
covariates $X_{F_i}$ as its input. Given individual estimators $f_i$
fixed, the problem of finding an optimal ensemble estimator of
additive form can be described as an optimization problem
\begin{equation}
 \min_{a_1,\ldots,a_D} \mathbb{E}\left[  \left( Y-\sum_{i=1}^{D}a_{i}f_i(X_{F_i})\right)
^{2}\right], \label{optproblem0}
\end{equation}
where $a_i$ are the weighting coefficients. Moreover, if we assume
that each estimator has no ``bias" after training, or equivalently,
if we assume that the residuals have zero mean, then it follows that
$\mathbb{E}[f_i(X_{F_i})]=\mathbb{E}[Y]$. Therefore, it is obvious
that the sum of all weighting coefficients is equal to $1$, i.e.
$\sum_{i=1}^D{a_{i}}=1.$

Consequently, we can rewrite the objective function as
\begin{equation}
\mathbb{E}\left[  \left(
\sum_{i=1}^{D}a_{i}\left[Y-f_i(X_{F_i})\right]\right) ^{2}\right].
\end{equation}
Note that the $i$th term in the parentheses is the residual of the
$i$th agent, defined as $R_i = Y - f_i(X_{F_i})$. Therefore, the
objective function can be rewritten as
\begin{equation}
\mathbb{E}\left[  \left( \sum_{i=1}^{D}a_{i}R_{i}\right)
^{2}\right].
\end{equation}

To simplify our derivation, define the covariance matrix of the
residuals as $\mathbf{A}$\, where $[\mathbf{A}]_{ij} =
\mathrm{cov}(R_i,R_j)$; then the problem can be further simplified
into a more concise form:
\begin{gather}
\min_{\mathbf{a}}\mathbf{a}^{T}\mathbf{Aa}\label{p1}\\
\text{s.t.}\text{ }\mathbf{1}^{T}\mathbf{a}=1 \label{p2}
\end{gather}
where $\mathbf{a}=\left[
\begin{array}
[c]{cccc}%
a_{1} & a_{2} & \cdots & a_{D}%
\end{array}
\right]  ^{T} $.
This is our starting point for the distributed
regression problem. The optimization problem of finding the best
ensemble estimator of the form of a linear combination of individual
estimators is equivalent to finding the best individual estimators
that generate the most desirable residuals that cancel each other
out.

The problem described by (\ref{p1}) and (\ref{p2}) is readily solved
if the covariance matrix $\mathbf{A}$\ is known and fixed. This is
equivalent to the problem of finding the best linear combination of
the estimators with these estimators given and fixed - a case that
appears in the non-cooperative training algorithms. However, in a
cooperative training algorithm, by communicating with each other,
the agents have a chance to change their training residuals
intelligently and repeatedly so as to reshape the covariance matrix
$\mathbf{A}$ and to minimize the ensemble error. It is this step
that makes the problem interesting and difficult. Therefore, the
entire problem can be summarized as a two-stage optimization
problem:
\begin{gather}
\min_{\mathbf{A}}\min_{\mathbf{a}}\mathbf{a}^{T}\mathbf{Aa}\label{stage2}\\
\text{s.t.}\text{ }\mathbf{1}^{T}\mathbf{a}=1\\
\mathbf{A} \text{ subject to training restrictions.}
\end{gather}
Note that $\mathbf{A}$ is subject to training restrictions because
the residual generated by each agent is not arbitrary. For $R_i$, it
must be achievable in the form of $Y - f_i(X_{F_i})$, which is
highly restrictive because of the space to which $f_i$ belongs.

\section{Solution to the two-stage optimization}
\noindent The first (inner) step of the optimization has a closed
form solution (solved by Lagrange multipliers). When
\begin{equation}
\mathbf{a}=\frac{\mathbf{A}^{-1}\mathbf{1}}{\mathbf{1}^{T}\mathbf{A}%
^{-1}\mathbf{1}},%
\end{equation}
the minimum value $\eta$ is achieved:
\begin{equation}
\eta=\frac{1}{\mathbf{1}^{T}\mathbf{A}^{-1}\mathbf{1}},
\end{equation}
i.e. the minimum value is the inverse of the sum of all the elements
of the inverse of $\mathbf{A}$. Moreover, since $\mathbf{A}$ is a
covariance matrix, and thus must be positive definite, the second
stage optimization problem is equivalent to
\begin{equation}
\max_{\mathbf{A}}\mathbf{1}^{T}\mathbf{A}^{-1}\mathbf{1}.
\label{opt2}
\end{equation}
It is necessary to bear in mind that $\mathbf{A}$ is subject to
training restrictions.

The optimization problem described in (\ref{opt2}) is the key step
of our algorithm. The most difficult step is to quantify the
``training constraints" of the covariance matrix of the residuals.
To tackle this problem, it is necessary to examine the inner
structure of $\mathbf{A}$.

As previously assumed, we have $D$ agents, and each agent $i$
maintains an estimator specified by the function $f_i(X_{F_i})$.
Then, obviously, the covariance matrix $\mathbf{A}$ can be expressed
as
\begin{equation}
[\mathbf{A}]_{ij} = \mathbb{E} \left[ \left( Y-f_i(X_{F_i}) \right)
\left( Y-f_j(X_{F_j}) \right)\right].
\end{equation}
where $[\mathbf{A}]_{ij}$ stands for the element of $\mathbf{A}$ in
the $i$th row and $j$th column. However, for numerical purposes, we
need to write the matrix in the form of a statistic by describing
everything in terms of actual data. So we characterize the function
$f_i$ by a vector $\mathbf{f}_i$, which is the prediction of the
function $f_i$ on all the training data points of agent $i$.
Similarly, we define $\mathbf{y}$ as the value of the outcome $Y$
for all data instances. Then, the covariance matrix $\mathbf{A}$ can
be estimated by (with the assumption of the unbiasedness of all the
estimators)
\begin{equation}
[\mathbf{A}]_{ij} = \frac{1}{N}\left[ \left(
\mathbf{y}-\mathbf{f}_i\right)^T \left(
\mathbf{y}-\mathbf{f}_j\right)\right].
\end{equation}

With this notation, the optimization problem can now be converted
into a more specific and implementable one:
\begin{gather}
\max_{\mathbf{f}_1 \ldots \mathbf{f}_D}\mathbf{1}^{T}\mathbf{A}^{-1}\mathbf{1}\\
\text{~~~~s.t.}\text{ }\mathbf{f}_i\in
\mathcal{H}_i,\text{}i=1,\ldots,D,
\end{gather}
where $\mathcal{H}_i$ denotes the space to which $\mathbf{f}_i$
belongs, which depends on the class of functions in which $f_i$
resides. Thus, the constraints on $\mathbf{A}$ are implicitly
included in the constraints of the vectors $\mathbf{f}_1,\ldots,
\mathbf{f}_m$.

The next step, ordinarily, is to massage the optimization problem
and to prove the convexity of the objective function and the domain
so that we can apply gradient descent algorithms and guarantee
global optimality. Yet for our problem, since the objective and
constraints are both rather intricate and to some extent only
implicitly specified, it is not very feasible to prove convexity
without additional assumptions. Therefore, we will directly develop
an algorithm based on gradient descent and test the algorithm
empirically before we delve deeply into the problem of global
optimality.

\subsection{Iterative covariance optimization algorithm}
\noindent
The first thing required for a gradient-based algorithm is
to find an expression for the gradient of the objective $\eta =
\mathbf{1}^T\mathbf{A}^{-1}\mathbf{1}$ with respect to
$\mathbf{f}_i$. By rather lengthy and intricate computation, a
closed form expression for the gradient is given by
\begin{eqnarray*}
\frac{\partial\eta}{\partial\mathbf{f}_{i}}&=&\frac{2}{|\mathbf{A}|^2}
\left( \mathbf{1}^T\mathbf{A}^{\ast}\mathbf{1}\right) \left(
\sum_{j=1}^{D}\left( \mathbf{y}-\mathbf{f}_{j}\right)
[\mathbf{A}^{\ast}]_{ij}\right)\\
&-&\frac{2}{|\mathbf{A}|}\left(\sum_{j\ne i}\left(
\mathbf{f}_{k}\mathbf{-f}_{j}\right)[\mathbf{B}
^{\ast}(k)]_{ij}\right), \label{grad1}
\end{eqnarray*}
where $k\ne i$, $\mathbf{A}^*$ denotes the adjoint of $\mathbf{A}$,
and $\mathbf{B}(k)$ is a $(D-1)\times(D-1)$ matrix given by
$$[\mathbf{B}(k)]_{ij}=(\mathbf{f}_k-\mathbf{f}_{i+\zeta_{ik}})^T(\mathbf{f}_k-\mathbf{f}_{j+\zeta_{jk}})$$
where $\zeta_{ik}=0$ if $i<k$, and $\zeta_{ik}=1$ if $i\ge k$.

This provides us with a feasible yet complex algorithm for
estimating the gradient. It is worth noting that the gradient
depends only on the residuals of the agents (through easy conversion
of $\mathbf{f}_i$ as $\mathbf{y}-\mathbf{r}_i$). In practice, we can
also use numerical methods to estimate the gradient, i.e. we can
perturb the components of $\mathbf{f}_i$, compute the change of the
objective and use the ratio between the change and the perturbation
as an approximation of that component of the gradient.

There is another important issue before we develop the algorithm: we
can search, by gradient descent, for a \emph{desirable}
$\mathbf{\hat{f}}_i$ to replace $\mathbf{f}_i$ so that we can
increase the value of the objective function, yet
$\mathbf{\hat{f}}_i$ might not be achievable because agent $i$ may
not be able to find a new estimator $\hat{f}_i$ such that
$\mathbf{\hat{f}}_i$ is realizable by $\hat{f}_i(X_{F_i})$.
Therefore, what is reasonable to do is to use $\mathbf{\hat{f}}_i$
as the new outcome for agent $i$ (instead of $\mathbf{y}$) to train
and find a new estimator $f_i$, i.e., we find the best projection of
$\mathbf{\hat{f}}_i$ onto the space $\mathcal{H}_i$.

Based on the description above, the basic idea of ICOA is summarized
as follows. First, cooperatively, all the agents determine the
present covariance matrix of their residuals $\mathbf{A}$. Then, one
by one, each agent finds its estimate of the gradient
$\frac{\partial\mathbf{1}^{T}\mathbf{A}^{-1}\mathbf{1}}{\partial\mathbf{f}_i}$,
after which the selected agent $i$ updates its vector $\mathbf{f}_i$
to $\hat{\mathbf{f}}_i $ using gradient descent. After that, agent
$i$ projects $\hat{\mathbf{f}}_i$ onto $\mathcal{H}_i$ by training
with $\hat{\mathbf{f}}_i$ as the outcome and thus obtains the new
version of $f_i$. Then, after agent $i$ updates its residual, all
the agents update their estimates of covariance matrix $\mathbf{A}$.

More precisely, the algorithm is as shown below:
~\\
~\\
\restylealgo{boxed}
\begin{algorithm}[H]
\SetLine

\While{$|\eta_n - \eta_{n-1}| > \epsilon$}{

    \For{$i$ from $1$ to $D$}{
    \begin{enumerate}
    \item Given current $\mathbf{A}$, compute $\frac{\partial \mathbf{1}^T\mathbf{A}^{-1}\mathbf{1}}{\partial
        \mathbf{f}_i}$;
    \item Back-search for the optimal step size $\Delta$;
    \item $\hat{\mathbf{f}}_i \leftarrow \mathbf{f}_i + \Delta \times
       \frac{\partial \mathbf{1}^T\mathbf{A}^{-1}\mathbf{1}}{\partial
       \mathbf{f}_i}$;
       \item Train $f_i(X_{F_i})$ with $\hat{\mathbf{f}}_i$ as the
        outcome;
        \item Use $f_i$ to update the training residual\\ of agent $i$ and update
        $\mathbf{A}$;
    \end{enumerate}
    }
    $\eta_{n+1} \leftarrow \mathbf{1}^T\mathbf{A}^{-1}\mathbf{1}$\;
    $n \leftarrow n + 1$\;
}
\end{algorithm}
~\\

\subsection{Simulation for regression problems}
\noindent In order to compare distributed regression implemented by
ICOA to other multi-dimensional regression algorithms (distributed
or non-distributed), we use three functions used in \cite{ref8} as
the hidden rule to generate our simulation training data sets. The
three functions and the corresponding joint distribution of the
covariates are:
\begin{itemize}
\item Friedman-1: $$\phi(\mathbf{x}) = 10\sin(\pi x_1 x_2)+20(x_3-1/2)^2+10x_4+5x_5+w,$$where
$x_j \sim U[0,1],\text{~}j=1,\ldots,5$;
\item Friedman-2: $$\phi(\mathbf{x}) = \left( x_1^2+\left( x_2x_3-\frac{1}{x_2x_4} \right)^2
\right)^{\frac{1}{2}}+w,$$ where
$$x_1\sim U[1,100],\text{~}x_2\sim U[40\pi,560\pi],$$
$$ x_3, x_5\sim U[0,1],\text{~}x_4\sim U[1,11].$$
\item Friedman-3: $$\phi(\mathbf{x}) = \tan^{-1}
\left(\frac{x_2x_3-\frac{1}{x_2x_4}}{x_1}\right)+w,$$ where the
distributions of the covariates are the same as those of Friedman-2.
\end{itemize}
\noindent All the covariates are independent of one another, and
before running the algorithm, the outcomes are normalized to the
range $[0,1]$. Also, to highlight the effects of the distributed
nature of the system, the independent additive white noise $w$ is
set to a negligible level in our simulation. Furthermore, it is
worth pointing out that in Friedman-2 and Friedman-3, attribute
$X_5$ is irrelevant, serving purely as a nuisance variable.

The structure of the entire distributed system is as follows. There
are 5 attributes, $X_1,\ldots,X_5$, and we assume that there are 5
agents, with agent $i$ observing attribute $X_i$ exclusively. Each
agent uses a regression tree as its individual estimator.

With the setup above, the simulation results of the two algorithms
are as shown in Table \ref{tab2}. As a comparison, we ran two other
distributed regression algorithms: averaging and residual refitting
(or ICEA, see \cite{c2} for details).

\begin{table}[!hbtp]
\centering
\begin{tabular}{|c|c|c|c|}
  \hline
  Friedman Data set & 1 & 2 & 3\\
  \hline
  ICOA & $.0047$ & $.0095$ & $.0086$ \\
  \hline
  Residual Refit & $.0047$  & $.0101$ & $.0096$\\
  \hline
  Averaging & $.0277$ & $.0355$ & $.0312$\\
  \hline
\end{tabular}
  \caption{Test errors (mean squared) of  ICOA, the residual refitting algorithm and the averaging algorithm on Friedman-1, -2 and -3.}
  \label{tab2}
\end{table}

\begin{figure}[h] 
  \centering
  \includegraphics[width = 40mm, height = 40mm]{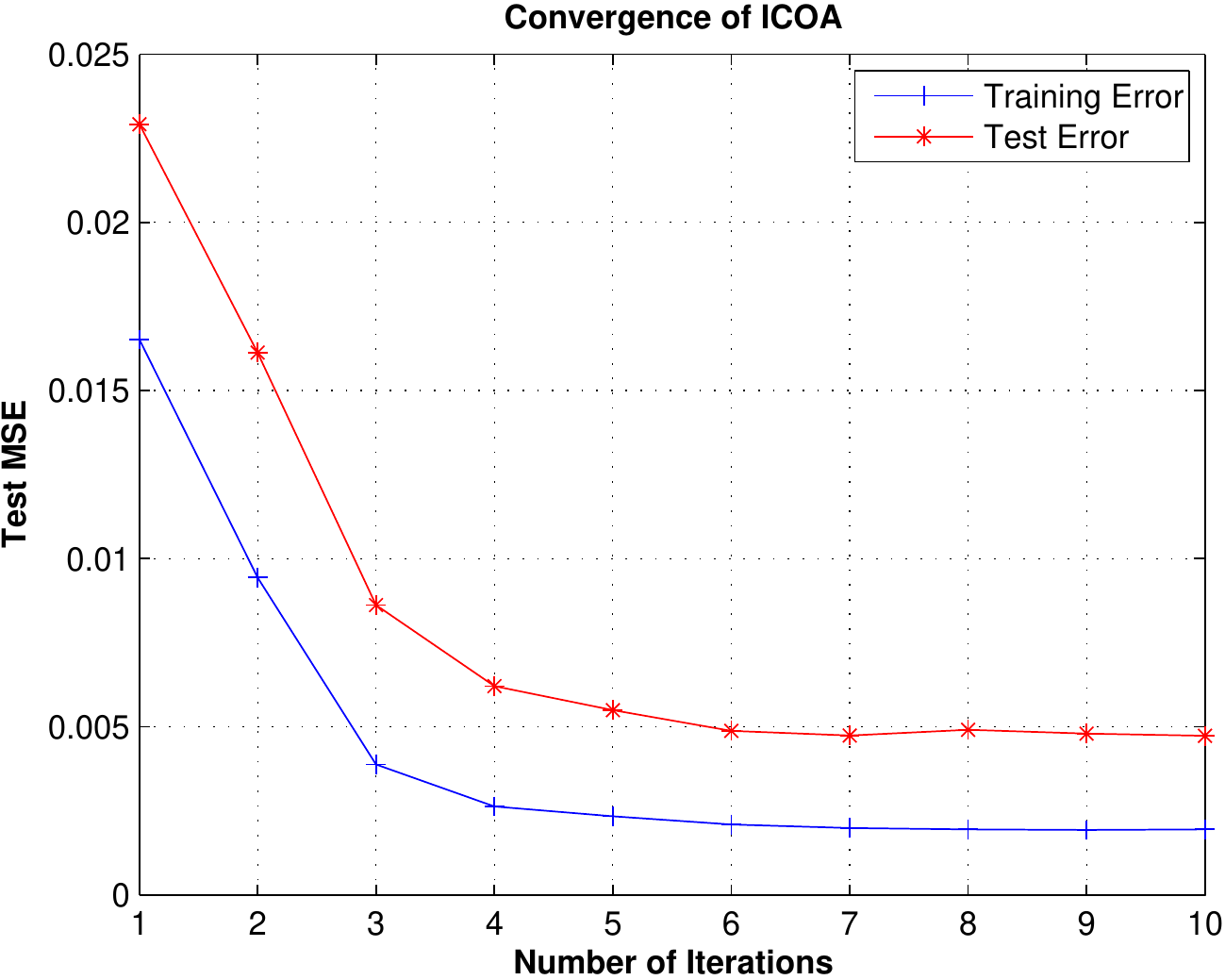}\,
  \includegraphics[width = 40mm, height = 40mm]{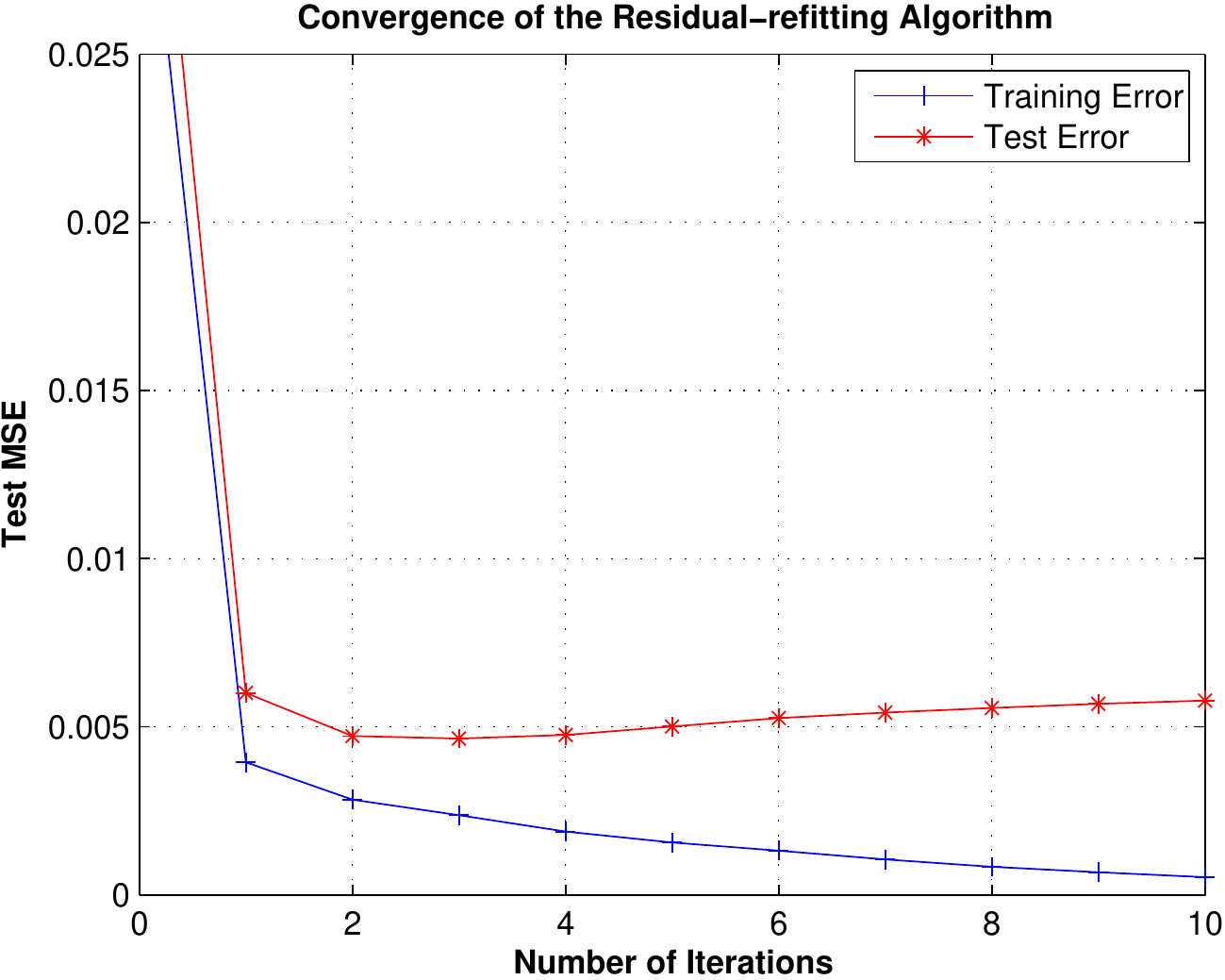}
  \caption{Comparison between the convergence of ICOA and the residual-refitting algorithm
  for Friedman-1. ICOA is less susceptible to
  overtraining. The training error of ICOA basically parallels the trends of the test
error. Yet for the residual-refitting algorithm, although the
training error converges to 0 rapidly, it does not correctly reflect
the trend of the test error of the ensemble estimator.}
  \label{overtrain}
\end{figure}

Generally speaking, the performance, measured in terms of test
error, of ICOA is slightly better than that of the residual
refitting algorithm for these three cases, while being much better
than the averaging algorithm. More importantly, ICOA has shown
little sign of overtraining, yet this is not the case for residual
refitting. This is demonstrated in Figure. \ref{overtrain}.

Note that for the residual refitting algorithm, the test error curve
turns up as the rounds of iteration increase, even if the training
error is consistently decreasing. On the contrary, the test error
curve and training error curve of ICOA are almost parallel and
horizontal. This suggests that ICOA knows when to stop unnecessary
overtraining and its training error is a good indicator of its test
error. This property is highly desirable.

The insusceptibility of ICOA to overtraining is a result of the fact
that whenever an agent optimizes its estimator, it takes into
consideration the predictions (represented by the residuals) of all
the other estimators, instead of only one (as in the case of the
residual refitting algorithm). Compared with residual refitting,
ICOA is a less ``greedy" algorithm in reducing its training error,
and once it reaches the optimal ensemble estimator, the covariance
matrix, known to all agents, would prevent the agents from changing
their estimators any further, unlike in the residual refitting
cases, where the agents are busy fitting the very last noise left in
the residuals, the culprit for overtraining.

\section{Optimization under inaccurate covariance}
\begin{figure}[!hbtp] 
  \centering
  \includegraphics[scale = 0.32]{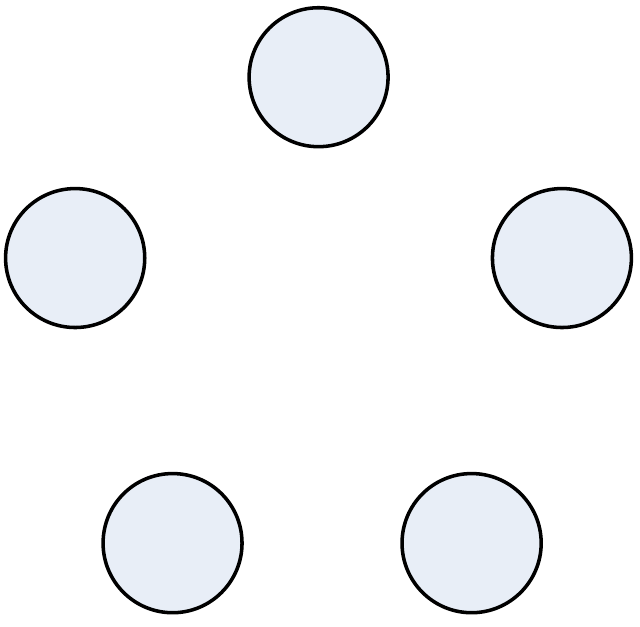}\,\,
  \includegraphics[scale = 0.32]{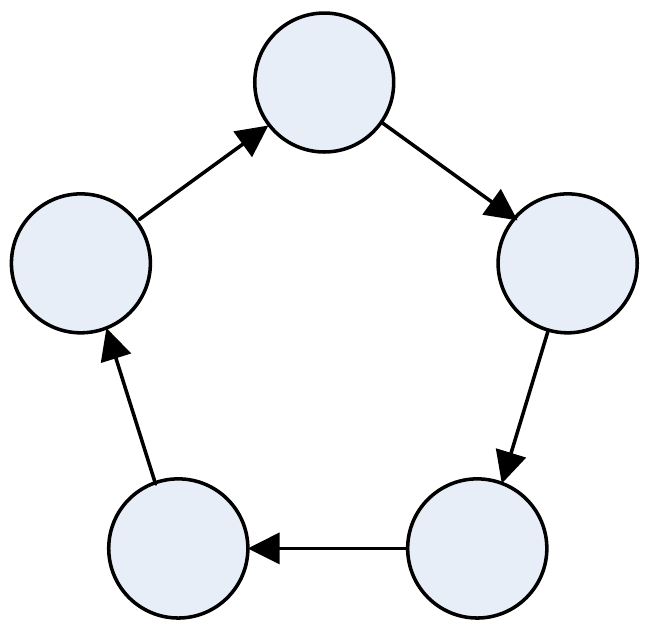}\,\,
  \includegraphics[scale = 0.32]{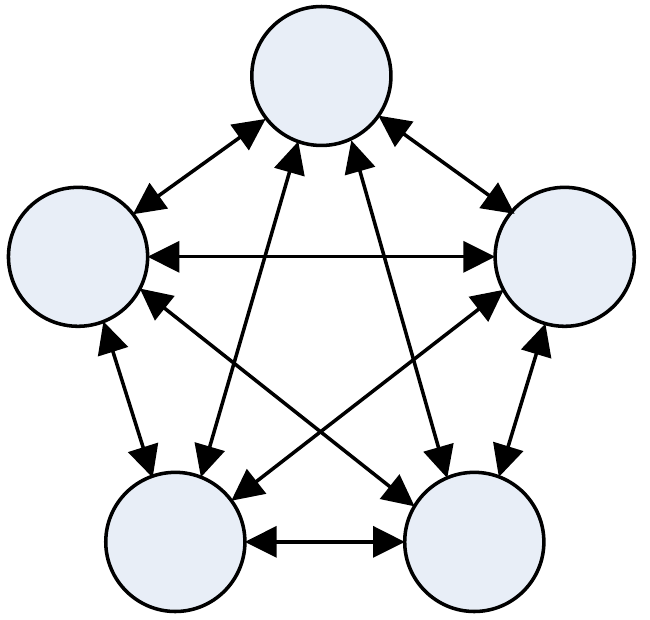}
  \caption{Illustrations of communication requirements for voting/averaging, residual refitting, and ICOA (from left to right).
   For a system of $D$ agents and $N$ training data instances, in terms of data transmission, the complexity for voting/averating algorithm is $O(1)$,
   the complexity for the residual refitting algorithm is $O(ND)$ and the complexity for ICOA is $O(ND^2)$.
  }
  \label{spacecomplexity}
\end{figure}

\noindent Although ICOA has an advantage in the performance of its
ensemble estimator compared to other distributed algorithms, it
requires more communication among the agents. In the
voting/averaging algorithm, no residual transmission is required. In
the residual refitting algorithm, for each iteration, residuals need
to be transmitted $D$ times in total, or once for each agent (as
soon as an agent finishes training, it sends its residual to the
next agent). However, for ICOA, residuals need to be transmitted
$D(D-1)$ times in total, or $D-1$ times for each agent (each agent
needs to send its residual to other agents whenever an agent
finishes training, because each agent needs all the latest training
residuals to compute the new covariance matrix). If the total number
of data instances is $N$, then in terms of data transmission, the
complexity is $O(1)$ for voting/averating, $O(ND)$ for the residual
refitting algorithm, and $O(ND^2)$ for ICOA. The most
communication-intensive algorithm among the three is ICOA. This is
highly undesirable when the number of agents is large. This is
illustrated in Figure \ref{spacecomplexity}.

To reduce the amount of data needed to be transmitted among the
agents for ICOA, we can relax the accuracy for the estimation of the
covariances. Yet this compromises the first (inner) step of our
two-stage optimization specified by (\ref{stage2}). We need to
develop an algorithm that still functions when $\mathbf{A}$ is not
fixed, but can take values over a non-singleton domain.

\subsection{A minimax problem}
\noindent Given the covariance matrix $\mathbf{A}$, the standard
optimization problem for ICOA is given by (\ref{p1}) and (\ref{p2}).
However, when we restrict the amount of data that can be
transmitted, the estimation of the covariance matrix is not accurate
enough. We can model this by allowing $\mathbf{A}$ to be of any
value in a range, i.e.
\begin{equation}
\mathbf{A}\in \mathcal{C}\cap\mathcal{P},
\end{equation}
where
\begin{equation}
\mathcal{C} = \{ \mathbf{S}\text{~}|\text{~}[\mathbf{S}]_{ij} \in
\left[\text{~}[\mathbf{A}_0]_{ij}-\delta_{ij},
[\mathbf{A}_0]_{ij}+\delta_{ij}\text{~}\right] \} \label{defC}
\end{equation}
and $\mathcal{P}$ stands for the class of semi-positive definite
matrices of the same size as $\mathbf{A}$. In other words, the
element $[\mathbf{A}]_{ij}$ has a range of length $2\delta_{ij}$
centered at $[\mathbf{A}_0]_{ij}$, with the semi-positivity of
$\mathbf{A}$ guaranteed.

This assumption addresses the inaccuracy of the estimation of the
covariance matrix $\mathbf{A}$. Moreover, the choice of $\mathbf{a}$
should take consideration of the worst case of $\mathbf{A}$, which
could be anywhere in $\mathcal{C}$ (here we neglect the extra
constraint of $\mathcal{P}$, and this makes our adversary
$\mathbf{A}$ even worse for the minimization), i.e., we have a
minimax optimization problem for the choice of $\mathbf{a}$:
\begin{gather}
\min_{\mathbf{a}}\max_{\mathbf{A}\in\mathcal{C}}\mathbf{a}^T\mathbf{A}\mathbf{a}
\label{minmax1}\\
\text{s.t. }\mathbf{a}^T\mathbf{1}=1.
\end{gather}
The solution to the inner maximization is straightforward, because
we can decompose this problem into $D\times D$ independent
optimization problems:
\begin{equation}
\max_{[\mathbf{A}]_{ij}}a_i a_j [\mathbf{A}]_{ij}.
\end{equation}
Obviously, when $a_i a_j
>0$, $[\mathbf{A}]_{ij}=[\mathbf{A}_0]_{ij}+\delta_{ij}$, otherwise,
$[\mathbf{A}]_{ij}=[\mathbf{A}_0]_{ij}-\delta_{ij}$. More concisely,
\begin{equation}
[\mathbf{A}]_{ij} = [\mathbf{A}_0]_{ij} + \delta_{ij} \sgn(a_i a_j).
\end{equation}

To simplify our next step, we now need to make a few more
assumptions. First, since the diagonal elements of $\mathbf{A}$ are
estimated locally, i.e., no data need be transmitted to estimate
residual variances of individual agents, it is reasonable to assume
$\delta_{ii}=0, \text{~}i=1,\ldots,D$. Another assumption is
$\delta_{ij}=\delta > 0, \text{~}i\ne j$; thus we can characterize
the uncertainty of the estimation of covariance using a single
number. This might sacrifice some accuracy of our model, yet it
simplifies our problem at least for a preliminary exploration.

With these additional assumptions, the optimal value $\zeta$ of the
maximization step in (\ref{minmax1}) is given by
\begin{equation}
\zeta = \mathbf{a} ^T\mathbf{A}_0\mathbf{a}+2\delta\sum_{i\ne
j}|a_i||a_j|.
\end{equation}
Thus, we can rewrite the objective of the minimax problem as
\begin{equation}
\min_{\mathbf{a}}\mathbf{a}
^T\mathbf{A}_0\mathbf{a}+2\delta\sum_{i\ne j}|a_i||a_j|\label{newmin}.\\
\end{equation}
Unfortunately, the objective function of this problem is not always
convex, and there is no closed form solution. To show the conditions
for convexity, we rewrite the objective function as
\begin{equation}
\mathbf{a} ^T\left( \mathbf{A}_0 - \delta \mathbf{I}\right)
\mathbf{a}+\delta(\sum_{i}|a_i|)^2.\label{innerMin}
\end{equation}
It is easy to show that the second term, the ``penalization term",
$\delta(\sum_{i}|a_i|)^2$, is a convex function. And the convexity
of the first term is dependent on the value of $\delta$. Since
$\mathbf{A}_0$ is a covariance matrix, i.e. it is positive definite,
the convexity of $\mathbf{a} ^T\left( \mathbf{A}_0 - \delta
\mathbf{I}\right) \mathbf{a}$ is hence equivalent to $\delta \le
\lambda_{\min}$, where $\lambda_{\min}$ is the smallest eigenvalue
of $\mathbf{A}_0$.

The second term serves as a penalization, restricting the magnitudes
of the coefficients. It is similar to Lasso Regression, except for
the square. This term can be crudely interpreted as follows: when
the covariance matrix is not accurately known, it is not wise to
fully minimize the ensemble training residual without paying
attention to the complexity of the ensemble model (measured by the
squared L-1 magnitude of the weighting coefficients).

Even if the problem is not convex, if the change in $\mathbf{A}$ is
not too large, the solution to (\ref{p1}) is a fairly good initial
value and gradient descent can be applied to solve the problem
specified by (\ref{newmin}) and (\ref{p2}).

\subsection{ICOA with Minimax Protection}
\noindent The above derivation actually changes the inner step of
our two-stage optimization, and we no longer have a closed form
solution. Nonetheless, we can still run ICOA numerically, because we
can still use perturbation to estimate the influence of the change
of $\mathbf{f}_i$ on the value of (\ref{innerMin}), given that we
can numerically solve the inner minimization.

Obviously, if we know the covariance accurately, changing the inner
step from minimization to minimax, i.e. changing (\ref{p1}) to
(\ref{newmin}) has no advantage. On the contrary, it compromises the
performance of the ensemble estimator and slows down the convergence
speed of ICOA. However, if we add restrictions on the number of data
instances exchanged between two agents, this makes $\mathbf{A}_0$,
the estimate of $\mathbf{A}$, less accurate, and then the minimax
optimization is of utmost importance for the convergence of ICOA. We
call this procedure \emph{Minimax Protection}.

\begin{figure}[!hbt] 
  \centering
  \includegraphics[scale = 0.4]{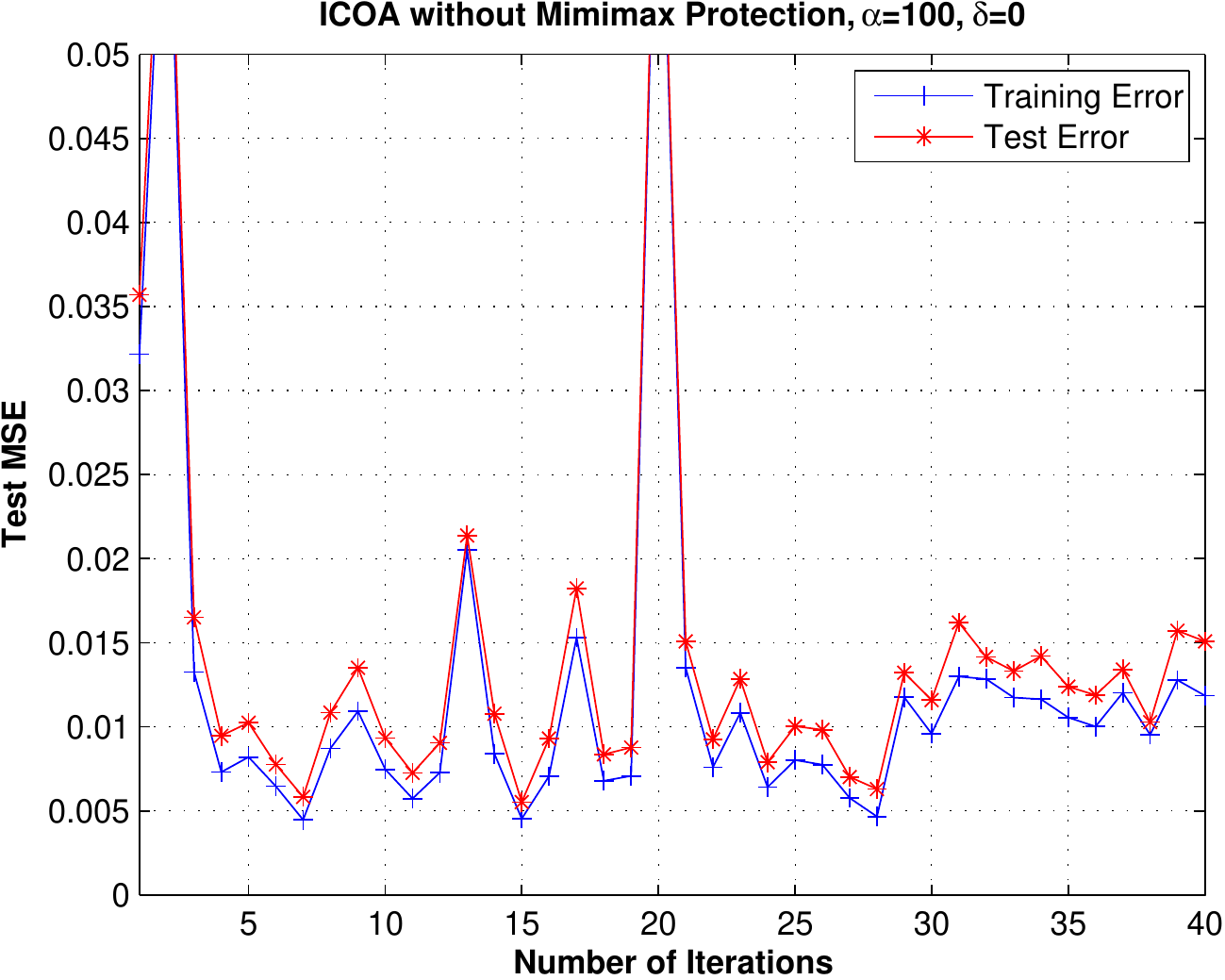}
  \caption{ICOA without Minimax Protection for Friedman-1. The training/test errors oscillates wildly and fail to converge. There is no way to decide when to stop the iterations.}
  \label{nocomp}
\end{figure}

For instance, if we transmit only $1/\alpha$ of the total $N$ data
instances (randomly sampled from all the data instances) for
covariance estimation, say, $\alpha = 100$, then the estimate
$\mathbf{A}_0$ has a large variance. Thus if we directly substitute
this estimate into the ICOA algorithm, it causes inaccurate and
unstable estimation of the direction for gradient descent and
prevents the algorithm from converging. This phenomenon is
illustrated in Figure \ref{nocomp}, where the compression rate is
$\alpha = 100$ (only $1\%$ of the data are transmitted for each
iteration) and $\delta = 0$ (no Minimax Protection).

\begin{figure}[!hbt] 
  \centering
  \includegraphics[scale = 0.4]{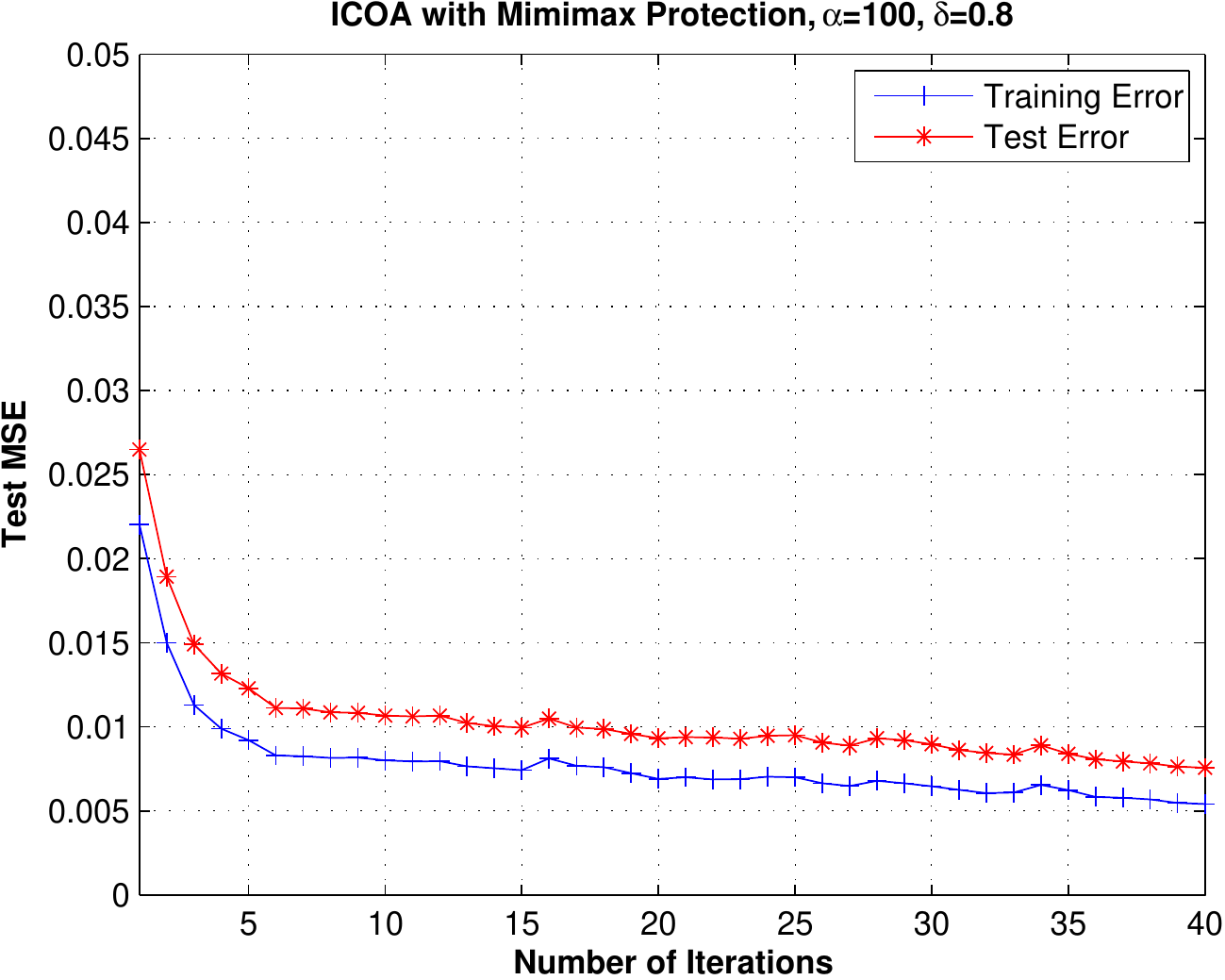}
  \caption{ICOA with Minimax Protection for Friedman-1. The training/test errors decrease almost monotonically and converge rather
  quickly and smoothly, with a reasonable compromise in performance.}
  \label{withcomp}
\end{figure}

However, if we foresee the inaccuracy and error of the estimation of
the covariance matrix, and properly choose a value for $\delta$,
then not only can we prevent the oscillation of the training/test
error, but we also sacrifice little in the performance of the
ensemble estimator. In Figure \ref{withcomp}, Minimax Protection is
applied to ICOA, with $\alpha = 100$ and $\delta = 0.8$.

The results of a series of simulations are shown in Table \ref{tab3}
for different values of compression rate $\alpha$ and $\delta$. In
this simulation, the data set is Friedman-1, and the system
configuration is the same as the simulation in the previous section.
The individual estimator is of the form of a $4$th order polynomial.
\begin{table}[!hbt]
\centering
\begin{tabular}{c|c|c|c|c|c}

  \hline
  \hline
    $\alpha$ & 1  & 10   & 50  &  200  & 800 \\
    \hline
$\delta = 0.00$  & .0037 & NaN &NaN &NaN &NaN\\
$\delta =.050$ & .0044  & .0045& NaN &NaN &NaN \\
$\delta =.500$ & .0051& .0056 &.0052& NaN& NaN \\
$\delta =.750$ & .0071 &.0071& .0073& .0077 &NaN  \\
$\delta =1.00$ &.0086 &.0086 &.0086 &.0090 &.0098 \\
$\delta = 2.00$ &.0112 &.0111 &.0112 &.0114 &.0113 \\
  \hline
  \hline
\end{tabular}
  \caption{Test errors (mean squared) of  ICOA with Minimax Protection for Friedman-1. For certain values of $\alpha$ and $\delta$, ICOA does not converge, and the test error exceeds machine limits and hence cannot be obtained.}
  \label{tab3}
\end{table}

It is worth pointing out two phenomena. First, when ICOA with
Minimax Protection converges, the performance is almost independent
of the compression rate $\alpha$. Second, given $\alpha$, when the
value of $\delta$ is above a certain level, ICOA almost always
converges, yet below that level, ICOA does not converge. These two
phenomena allow us to find an optimal $\delta$ for every given
$\alpha$, so that we can optimize the performance of ICOA under a
given compression rate.

In Table \ref{tab3}, another dramatic phenomenon is the case for
$\alpha = 800$ and $\delta = 1.00$. Since we have only $4000$
training data instances, this means in each iteration, we use only
$5$ pairs of numbers to estimate the covariance between two agents.
And Minimax Protection with properly selected $\delta$, enables us
to achieve a decent test mean square error of $.0098$, only about
$2.5$ times of the optimal value $.0037$. Yet only $1\%$ of the data
transmission is needed compared with the amount needed in the
optimal case (after taking into consideration the longer convergence
time). Thus ICOA provides us with a very useful tool to trade off
between performance and data transmission.

\subsection{Upper bound of the test error}
\noindent From the simulation results shown in Table \ref{tab3}, it
is of interest to investigate the relationship between the
compression rate $\alpha$ and the optimal performance (measured by
test error) of the system. As analyzed previously, the key is to
select a proper $\delta$ so that we neither under-protect ICOA
(leading to unstable convergence) nor over-protect (leading to worse
performance). This requires us to investigate the statistical
properties of the estimator of the correlation coefficient between
two random variables. In \cite{ref9}, it is shown that the pivot
statistic $T_N$ of the sample correlation coefficient has the
Student's t-distribution; that is,
\begin{equation}
T_N =  \frac{\sqrt{N-2}\hat{\rho}}{\sqrt{1-\hat{\rho}^2}} \sim
t_{N-2},
\end{equation}
where $N$ is the number of data instances. Therefore the $95\%$
confidence interval of the correlation coefficient is given by
$[\hat{\rho}-\xi, \hat{\rho}+\xi]$, where $\xi
=1.96(1-\hat{\rho}^2)/\sqrt{N}$.

If we assume that the largest variance of all the residuals is
$\sigma_{\max}^2$, then an approximation to the optimal $\delta$ (as
a function of $\alpha$) can be given by
\begin{equation}
\delta_{\mathrm{opt}}(\alpha)=\min\{1.96\sigma_{\max}^2/\sqrt{N/\alpha},2\sigma_{\max}^2\}.
\label{optdelta}
\end{equation}
The basic idea is to find the smallest $\delta$ that covers, with
high probability, the possible domain of the covariance matrix,
given a crude estimate $\mathbf{A}_0$.

\begin{figure}[!hbt] 
  \centering
  \includegraphics[scale = 0.45]{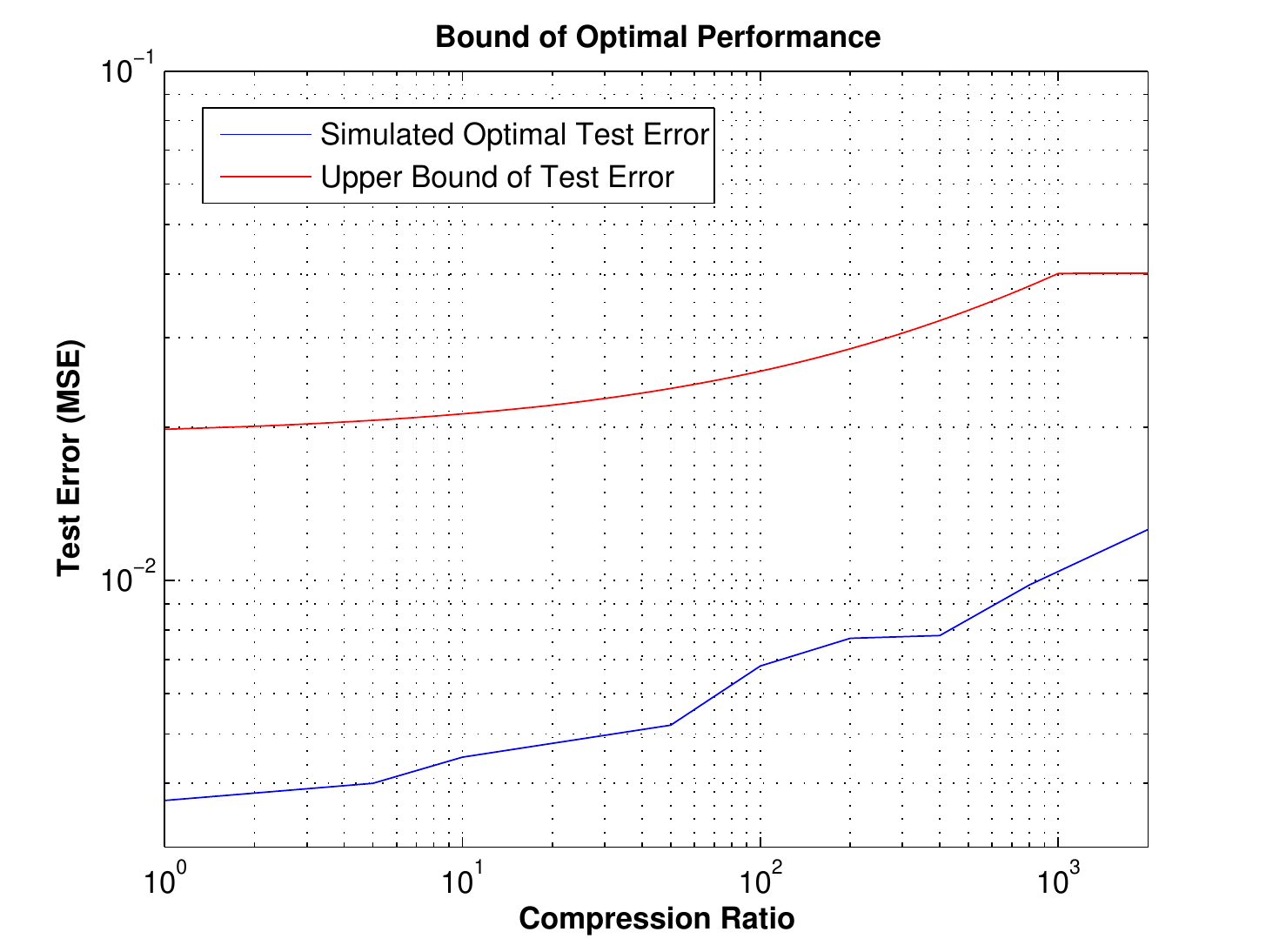}
  \caption{Comparison between the upper bound given by (\ref{ub}) and the simulated test error.}
  \label{upperbound}
\end{figure}

With this approximation, we are able to develop an upper bound on
the test mean square error as a function of $\alpha$. Define
$\mathbf{A}_{\mathrm{ini}}$ as the covariance matrix (accurate) of
the residuals of all individual estimators before we run ICOA. For
each step, ICOA with Minimax Protection improves the test error (not
merely training error), because Minimax Protection guarantees, with
high probability, that the true covariance matrix is in the range
$\mathcal{C}$ defined in (\ref{defC}). Therefore, the solution to
\begin{equation}
\min_{\mathbf{a}}\mathbf{a}^T(\mathbf{A}_{\mathrm{ini}}-\delta_{\mathrm{opt}}(\alpha)\mathbf{I})\mathbf{a}+\delta_{\mathrm{opt}}(\alpha)(\sum_{i}|a_i|)^2
\label{ub}
\end{equation}
with constraint (\ref{p2}) provides us with an upper bound (with
high probability) on the generalization error with respect to the
compression rate $\alpha$. Figure \ref{upperbound} illustrates the
comparison between this upper bound and the simulated optimal
performance.

\section{Conclusions}
\noindent In this paper, we have shown that ICOA, as a cooperative
training algorithm, demonstrates its efficacy for finding an optimal
ensemble estimator of additive form, while demonstrating an
insusceptibility to overtraining. Moreover, Minimax Protection
provides us with a tool to run ICOA when covariances are not
accurately estimated, and hence enable us to trade off between
performance and data transmission. Minimax Protection, combined with
ICOA, also helps us to develop an upper bound on the test error for
the ensemble estimator.

\end{document}